\documentclass[]{emulateapj}

\shorttitle{High Energy Gamma-ray Absorption and Cascade Emission in Nearby Starburst Galaxies}
\shortauthors{Yoshiyuki Inoue}

\usepackage{apjfonts}
\begin{document}


\title{High Energy Gamma-ray Absorption and Cascade Emission in Nearby Starburst Galaxies}

\author{Yoshiyuki Inoue}
\affil{Department of Astronomy, Kyoto University, Kitashirakawa, Sakyo-ku, Kyoto 606-8502, Japan}

\email{yinoue@kusastro.kyoto-u.ac.jp}

\begin{abstract}
High energy gamma-ray emission from two nearby bright starburst galaxies, M82 and NGC 253, have recently been detected by {\it Fermi}, H.E.S.S., and VERITAS. Since starburst galaxies have a high star formation rate and plenty of dust in the central starburst region, infrared emissions are strong there. Gamma-ray photons are absorbed by the interstellar radiation field photons via electron and positron pair creation. The generated electron and positron pairs up scatter the interstellar photons to very high energy gamma-ray photons via cascade emission through inverse Compton scattering. In this paper, we evaluate the contribution of this cascade emission to the gamma-ray spectra of M82 and NGC 253. Although it would be difficult to see direct gamma-ray evidence of cosmic-rays with an energy $>10$ TeV due to the gamma-ray attenuation, the resulting cascade emission would be indirect evidence. By including the cascade component, we find that the total flux above 1 TeV  increases $\sim$ 18\% and $\sim$ 45\% compared with the absorbed flux assuming the maximum kinetic proton energy as 45.3 TeV and 512 TeV, respectively. Future gamma-ray observatories such as CTA would be able to see the indirect evidence of cosmic-ray with an energy $>10$ TeV by comparing with theoretical emission models including this cascade effect.
\end{abstract}



\section{Introduction}
Supernova remnant (SNR) shocks are believed to accelerate cosmic-rays \citep{gin64,hay69}. Although this was confirmed by the detection of non-thermal emissions from the Galactic SNRs in X-ray and $\gamma$-rays \citep{koy95,aha04}, the cause of the highest energy galactic cosmic-rays is still under discussion. The highest energy of the Galactic cosmic-ray are thought to be comparable with the {\it knee} energy of the cosmic-ray spectrum, $\sim10^{15.5}$ eV \citep{nag00}. Very high energy (VHE; $>$30 GeV) $\gamma$-ray observations of SNRs shows that SNRs have a cut-off around 10 TeV in the $\gamma$-ray spectra or very soft injection proton spectra \citep[see e.g.][]{abd10_casa}. Thus, it is not possible to simply expect that the SNR shocks accelerate cosmic-rays up to the energy around the {\it knee} energy region. Further investigation about these problems is currently required.

Since starburst galaxies have a high star formation rate (i.e. high supernova rate) and plenty of gas, they are a candidate for proving the cosmic-ray acceleration scenario as well as the Galactic SNRs. Many papers have studied the $\gamma$-ray emission mechanism in starburst galaxies \citep{voe89,aky91,pag96,tor04,ds05,per08,dcdp09,rep10}. Very recently, high energy $\gamma$-ray emissions from the nearby bright starburst galaxies, M82 and NGC 253, have been detected by {\it Fermi}, H.E.S.S., and VERITAS \citep{abd10_sb, ace09, ver09}. These detections support the idea of $\gamma$-rays generated by cosmic-ray propagation in starburst galaxies. Observationally, however, the existence of high energy emissions $>10$ TeV has not been satisfactorily constrained yet, although it would be a probe of the existence of cosmic-rays with an energy $>10$ TeV.

It is well known that VHE $\gamma$-rays propagating through the universe are absorbed by the cosmic optical-infrared background radiation via positron and electron, $e^+e^-$, pair production \citep{gou66,jel66}. This $\gamma$-ray attenuation process also occurs in starburst galaxies which have plenty of infrared photons \citep{ds05,dcdp09}. It might be difficult to detect $>10$ TeV emissions from starburst galaxies due to this attenuation process.

Here, created electron-positron pairs by attenuation up-scatter the interstellar photons to VHE $\gamma$-ray photons via the inverse Compton scattering process. This is the so called cascade emission \citep{aha94,wan01,dai02,raz04,and04,mur07,kne08,ino09,ven10}. If the energy of the intrinsic $\gamma$-rays extends to $>$10 TeV, such a cascade emission would appear in the VHE band. This cascade emission would be indirect evidence of the existence of intrinsic $>10$ TeV emissions. Therefore, we need to take into account cascade emissions to investigate the existence of cosmic-ray with an energy $>10$ TeV. In the past studies on starburst galaxies, however, the effect of cascade emission was not taken in to account.

In this paper, we estimate the contribution of the cascade emission to the $\gamma$-ray spectra for M82 and NGC 523. To investigate this effect, we calculate the electron-positron pair creation optical depths in these starburst galaxies. We also compare the expected spectra including absorption and cascade effects with observed $\gamma$-ray spectra. 

\section{Absorption and Cascade Emission Modeling}
\label{sec:model}

It is well known that VHE photons from high redshifts are absorbed by the interaction with the intergalactic optical-infrared radiation field via electron-positron pair creation \citep{gou66,jel66}. It is also theoretically expected that these created pairs would scatter the cosmic microwave background radiation as secondary $\gamma$-ray emissions (the so called cascade emission) in the case of blazars, gamma-ray bursts, and the extragalactic $\gamma$-ray background radiation \citep{aha94,wan01,dai02,raz04,and04,mur07,kne08,ino09,ven10}. We investigate this cascade emission effect in two nearby starburst galaxies, M82 and NGC 253. 

\subsection{Absorption of VHE $\gamma$-ray photons in starburst galaxies}
\label{subsec:abs}

 \begin{figure*}
  \begin{center}
\centering
\epsscale{1.1}
\plottwo{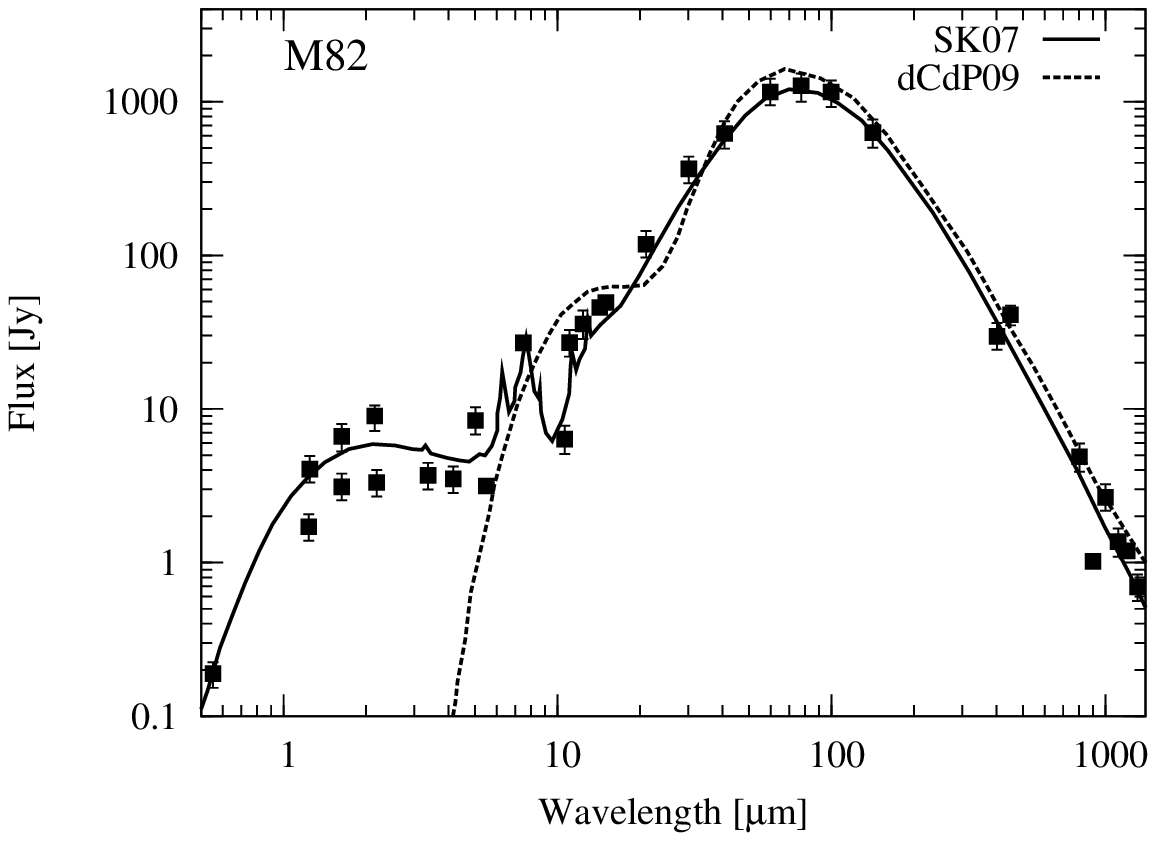}{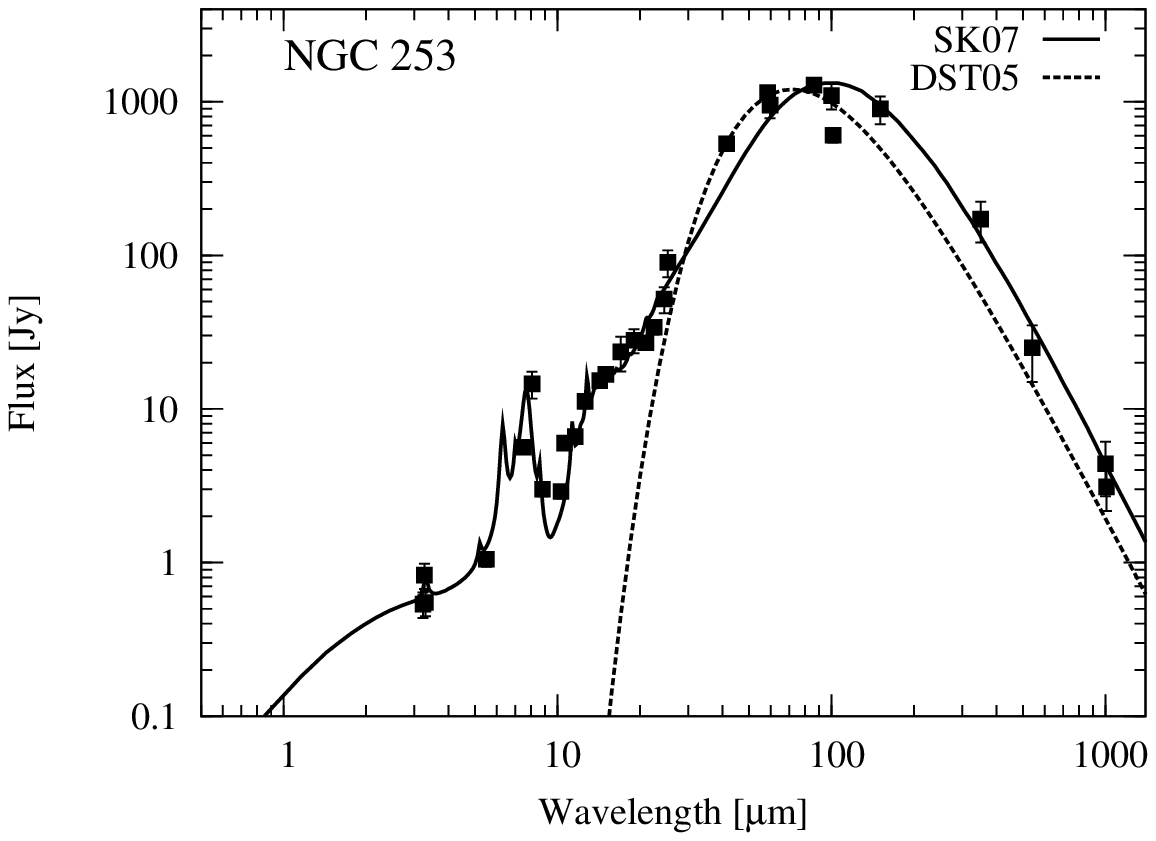}
\caption{{\it Left} panel: SED of M82. The solid and dashed curve represents \citet[][SK07]{sie07} and \citet[][dCdP09]{dcdp09} SED model, respectively. Data references (near infrared: \citet{joh66,kle70,aar77,jar03}; far infrared: \citet{rie72,tel80,rie80,kle88,tel92,for03}; submm: \citet{jaf84,kru90,hug90,hug94}). {\it Right} panel: SED of NGC 253. The solid and dashed curve represents SK07 and \citet[][DST05]{ds05} SED model, respectively. Data references (near infrared: \citet{rie75,rad01}; far infrared: \citet{tel80,mel02}; submm: \citet{rie73,hil77,eli78,chi84}).}
\label{fig:SED}
\end{center}
\end{figure*}

 Very recently two nearby bright starburst galaxies, M82 and NGC 253, have been detected with $\gamma$-ray observations by {\it Fermi}, H.E.S.S., and VERITAS \citep{abd10_sb,ace09,ver09}. These $\gamma$-ray detected starburst galaxies would not suffer from $\gamma$-ray absorption with the intergalactic radiation field, since they are very close to us. The distance to M82 and NGC 253 is $3.6\pm0.3$ Mpc \citep{fre94} and $3.9\pm0.4$ Mpc \citep{kar03}, respectively. The starburst region is located in the inner part of the galaxy with a size of 500 pc and 280 pc for M82 and NGC 253, respectively \citep{may06,ulv00}. 

 The central starburst region in these galaxies where SNRs are produced is theoretically expected to emit most of the $\gamma$-ray photons ({Domingo-Santamar{\'{\i}}a} \& {Torres} 2005, hereinafter DST05; {de Cea del Pozo} {et~al.} 2009, hereinafter dCdP09). In the case of the Large Magellanic Cloud, the association between starburst region and $\gamma$-ray emission region is observationally confirmed by {\it Fermi} \citep{abd10_lmc}. It is also known that there are plenty of interstellar radiation field photons from star forming activities. Therefore, we need to take into account the $\gamma$-ray absorption effect in the central star formation region.

\begin{figure}
  \begin{center}
\centering
\plotone{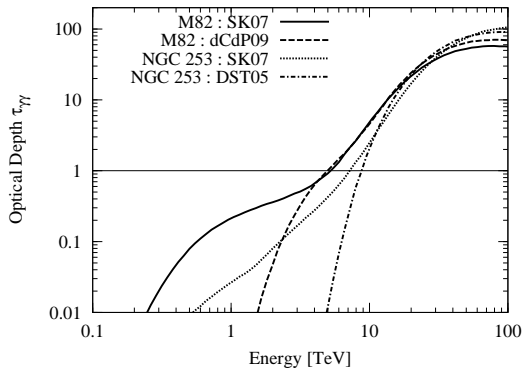}
\caption{The optical depth of pair production, $\tau_{\gamma\gamma}$, in starburst galaxies as a function of the photon energy. Solid, dashed, dotted and dot-dashed curve corresponds to $\tau_{\gamma\gamma}$ in SK07 model for M82, dCdP09 model for M82, SK07 model for NGC 253 and DST05 model for NGC 253, respectively. The horizontal line marks the level of optical depth $\tau_{\gamma\gamma}=1$.}
\label{fig:tau}
\end{center}
\end{figure}

\citet[][hereinafter SK07]{sie07} developed optical-infrared spectral energy distribution (SED) models for the nuclei of starburst galaxies and ultra luminous infrared galaxies by computing a radiative transfer SED of spherical, dusty galactic nuclei. They include silicates, amorphous carbon, graphite grains, and polycyclic aromatic hydrocarbons in the form of dust. Fig. \ref{fig:SED} shows the optical-infrared spectral energy distribution (SED) of M82 and NGC 253. We adopt their spectrum model for M82 and NGC 253 shown in Fig. 2 and 3  in their paper, which is consistent with the data. In the case of M82, we add a blackbody component ($T=2500$ K) to fit the data below 5 $\mu {\rm m}$ as in \citet{sie07}, which was not taken into account in previous works. For comparison, we also show the interstellar SED model in dCdP09 and DST05 for M82 and NGC 253, respectively. Since SK07 models well reproduce the observed data over a wide range of wavelengths for both galaxies, hereafter we use the SK07 optical-infrared interstellar SED models as our baseline nuclei SED models.

Fig. \ref{fig:tau} shows the expected optical depth, $\tau_{\gamma\gamma}$, of M82 for SK07 and dCdP09 models and NGC 253 for SK07 and DST05 models. We do not take into account the interaction of a $\gamma$-ray with a nuclei, since this would not absorb $\gamma$-rays significantly \citep{ds05}. The $\gamma-\gamma$ absorption becomes significant (i.e. $\tau_{\gamma\gamma}=1$) above $\sim 5$ TeV and $\sim9$ TeV for M82 and NGC 253, respectively. 

\subsection{Calculation of VHE $\gamma$-ray spectra}
\label{subsec:spe}

The $\gamma$-ray emissions from starburst galaxies are modeled by many papers \citep{voe89,aky91,pag96,tor04,ds05,tho07,per08,dcdp09,rep10}. In this paper, we substitute the $\pi_0$-meson decay spectrum for the intrinsic $\gamma$-ray spectrum, $dN_i/dE_\gamma$, using code provided by \citet{kar08} \citep[see also][]{kam05,kam06}. For the total inclusive inelastic {\it p}-{\it p} cross section, they include the non-diffractive (with Feymann scaling violation) and diffractive components, plus the $\Delta$(1232) and Res(1600) resonance excitation contributions. The normalization of the total $\gamma$-ray flux including absorption and cascade effects is adjusted to the observed {\it Fermi} photon flux data in the 0.1--5 GeV band. The intrinsic $\gamma$-ray spectrum is given by the highest proton kinetic energy, $T_{p, \rm max}$, in units of [TeV] and the spectral index of the intrinsic proton spectrum, $\Gamma$, where we set $dN_p/d\gamma_p\propto \gamma_p^{-\Gamma}$, $dN_p/d\gamma_p$ the proton flux, $\gamma_p$ the Lorentz factor of protons. We do not take into account inverse Compton scattering and bremsstrahlung radiation, since their contribution is expected to be less than 10 \% above 0.1 GeV (DST05, dCdP09). Although we need to solve the cosmic-ray propagation in starburst galaxies for detailed calculations, we adopt a simple analytical model to see the absorption and cascade effects.

 \begin{figure*}
  \begin{center}
\centering
\epsscale{1.1}
\plottwo{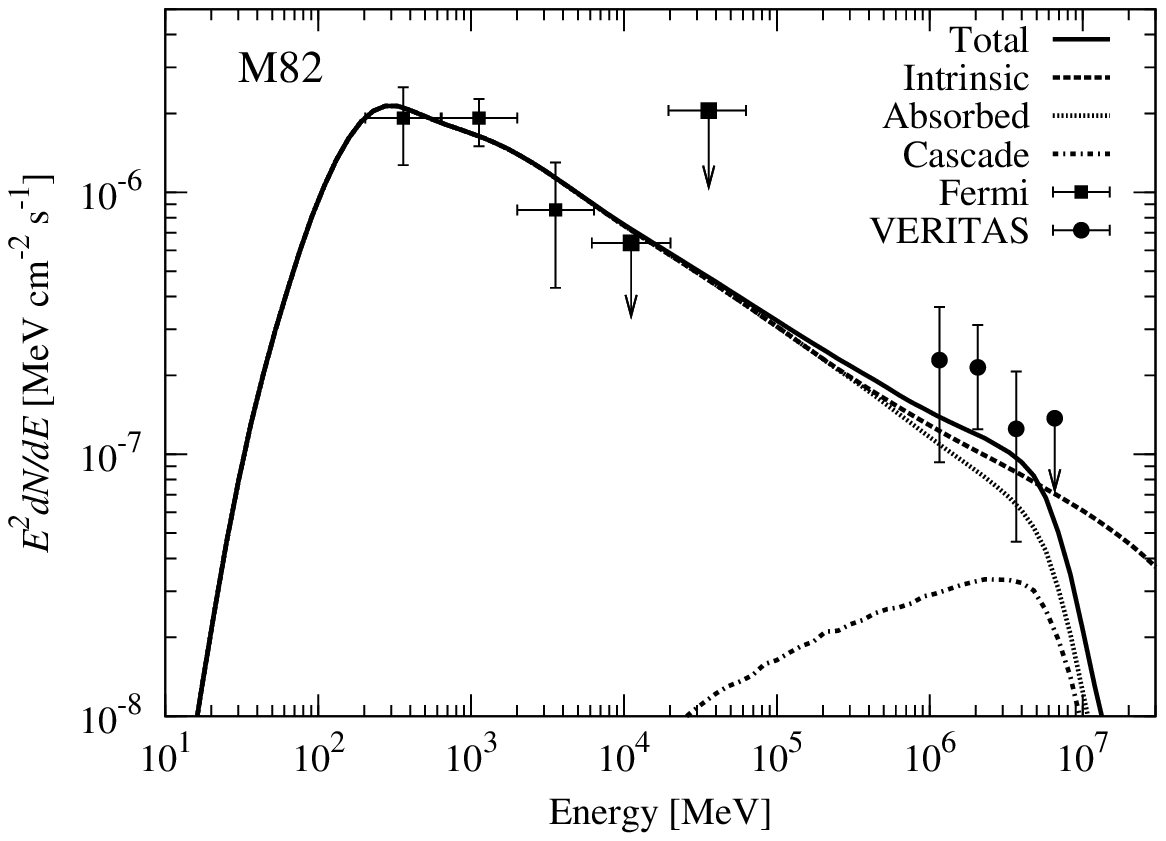}{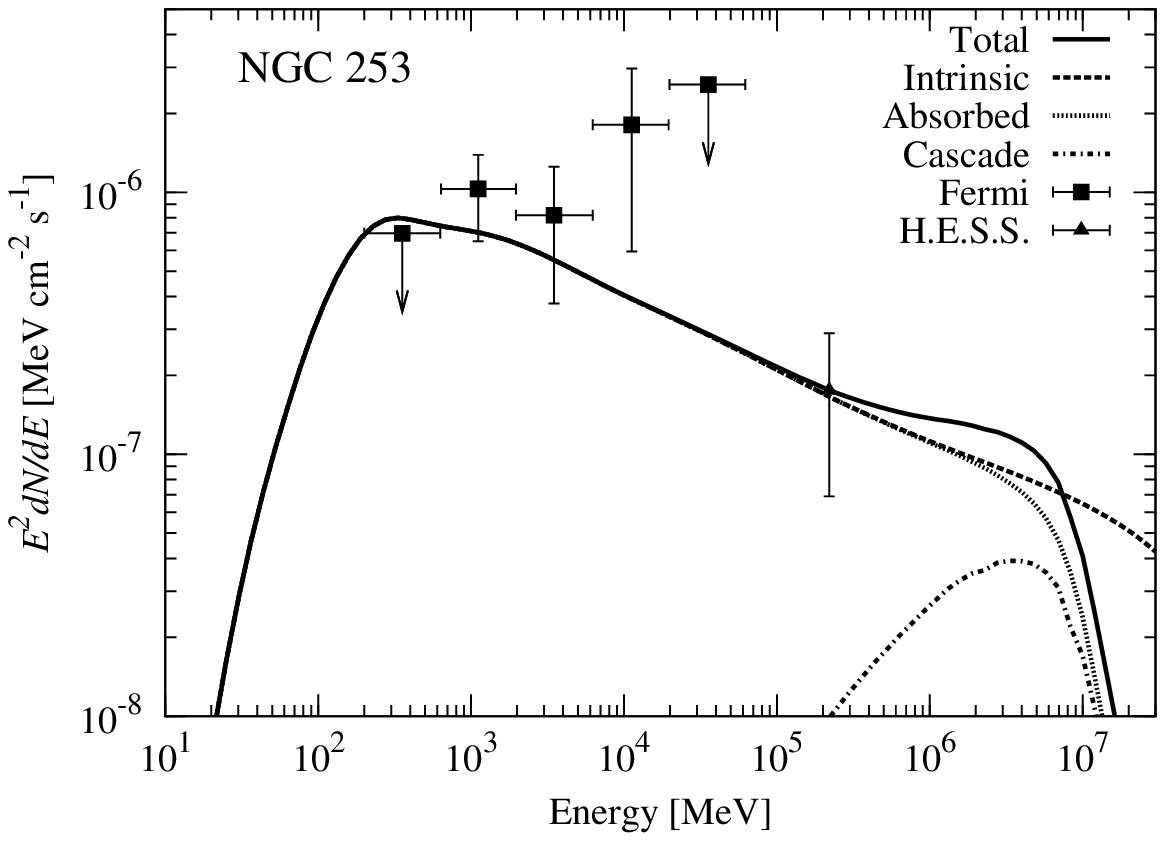}
\caption{{\it Left} panel: Gamma-ray spectrum of M82. The solid curve is the total flux, i.e. absorbed plus cascade component. Three model predictions for the intrinsic (no absorption), absorbed, and cascade components are also shown by the curve marking shown in the figure. We adopt ($\Gamma, T_{p, \rm max}) = (2.4, 512\  {\rm TeV})$ for the intrinsic spectrum. For interstellar radiation field SED model, we adopt the SK07 model. The observed data of {\it Fermi} \citep{abd10_sb} and VERITAS \citep{ver09} are also shown. {\it Right} panel: Same as the {\it left} panel, but for NGC 253 with ($\Gamma, T_{p, \rm max}) = (2.3, 512 \ {\rm TeV})$. The observed data of {\it Fermi} \citep{abd10_sb} is also shown. The single VHE flux point for NGC 253 is computed from the integral photon flux $>$220 GeV reported by the H.E.S.S. \citep{ace09} assuming a power-law spectral model with photon index between 2.0 and 3.0.}
\label{fig:f_sb}
\end{center}
\end{figure*}

For the calculation of the cascade emission, we ignore magnetic fields, since the $\gamma$-ray emissions from starburst galaxies are isotropic and continuous. We calculate the resulting cascade emissions as follows:

\begin{equation}
\frac{dN_c}{dE_\gamma}=\int_{\gamma_{e,\rm min}}^{\gamma_{e,\rm max}} d\gamma_e\frac{dN_e}{d\gamma_e}\frac{dN_{\gamma_e,\epsilon}}{dtdE_{\gamma,i}}t_{IC}
\label{eq:cascade}
\end{equation}
\citep{blu70,aha94,wan01,dai02,raz04,and04,mur07,kne08,ino09,ven10}, where $t_{IC}$ is the inverse Compton (IC) cooling time scale and the electron injection spectrum is
\begin{equation}
\frac{dN_e}{d\gamma_e} = 2 \frac{dE_{\gamma,i}}{d\gamma_e}\frac{dN_i}{dE_{\gamma,i}} (1 - e^{-\tau_{\gamma,\gamma}(E_{\gamma,i})} )
\end{equation}

and the scattered photon spectrum per unit time by IC scattering is:

\begin{equation}
\frac{dN_{\gamma_e,\epsilon}}{dtdE_{\gamma,i}}=\frac{2\pi r_0^2 c}{\gamma_e^2}\int d\epsilon \frac{1}{\epsilon}\frac{dn}{d\epsilon}(\epsilon) f(x)
\end{equation}
with $f(x)=2x \ln(x)+x+1-2x^2$, ($0<x<1$) and $x = E_{\gamma,i}/4\gamma_e^2\epsilon$. Here, $E_{\gamma,i}=2\gamma_e m_e c^2$ is the energy of intrinsic photons, ${dN_i}/{dE_{\gamma,i}}$ is the intrinsic $\gamma$-ray spectrum, $r_0$ the classical electron radius, $dn/d\epsilon(\epsilon)$ is the photon density at the starburst region. The integration region over the Lorentz factor, $\gamma_e$, is $\gamma_{e,{\rm min}}<\gamma_e<\gamma_{e,{\rm max}}$, $\gamma_{e,{\rm max}}=E_{\gamma,{\rm max}}/2$ and $\gamma_{e,{\rm min}}={\rm max}[m_ec^2/2\epsilon,(E_\gamma/\epsilon)^{1/2}/2]$, where $E_{\gamma,{\rm max}}$ is the maximum energy of $\gamma$-ray photons.

We iteratively calculate Eq. \ref{eq:cascade} by substituting $dN_c/dE_\gamma (1-e^{-\tau_{\gamma,\gamma}})$ in order to include IC scattering due to generated pairs from reabsorbed secondary photons.

\section{Results} 

 Fig. \ref{fig:f_sb} shows the $\gamma$-ray spectra of M82 and NGC 253 in units of [MeV/cm$^2$/s]. We show the intrinsic (the spectrum without taking into account absorption), absorbed, and cascade components as well as total (absorbed + cascade) spectra. For the intrinsic spectra, we adopt ($\Gamma, T_{p, \rm max}) = (2.4, 512\  {\rm TeV})$ and $(2.3, 512 \ {\rm TeV})$ for M82 and NGC 253 as our standard models, respectively. In the case of M82, our $\gamma$-ray spectrum nicely fits to the observed data in the range of the uncertainty of the data by taking into account the $\gamma-\gamma$ absorption and the cascade emissions. However, in the case of NGC 253, it is difficult to explain the {\it Fermi} data point at 10 GeV, even if we take into account the secondary $\gamma$-ray populations. Since the absorption is effective above 10 TeV and the seed photon density has a peak around 0.01 eV, the typical energy of the secondary $\gamma$-ray, $E_c$, is $E_c\sim(10 {\rm TeV}/2m_ec^2)^2\times0.01 {\rm eV}\sim1 {\rm TeV}$. Thus, cascade emission do not contribute at $\sim$10 GeV, rather, closer to around $\sim 1$ TeV. By including cascade emissions, the total $\gamma$-ray flux above 1 TeV becomes 43\% and 46\% greater than the absorbed flux for M82 and NGC 253, respectively. 
   
\begin{figure*}
  \begin{center}
\centering
\epsscale{1.1}
\plottwo{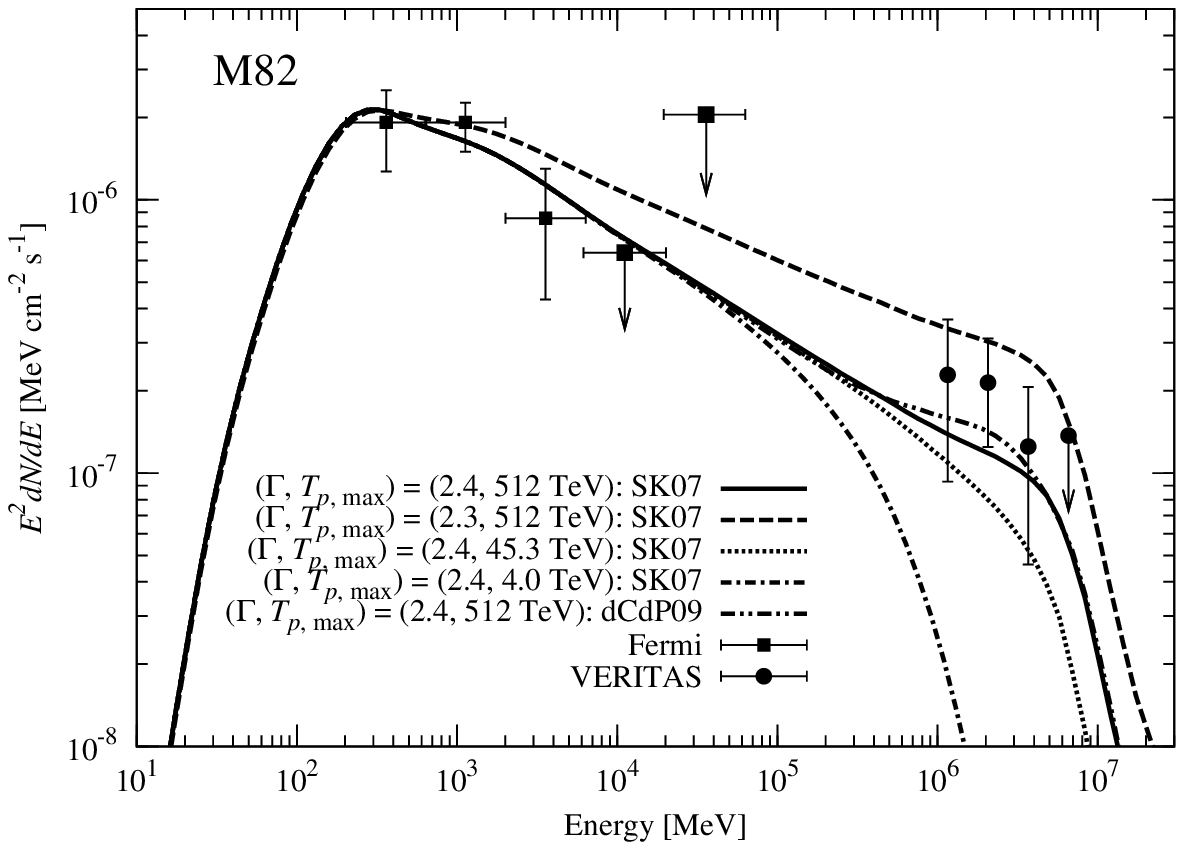}{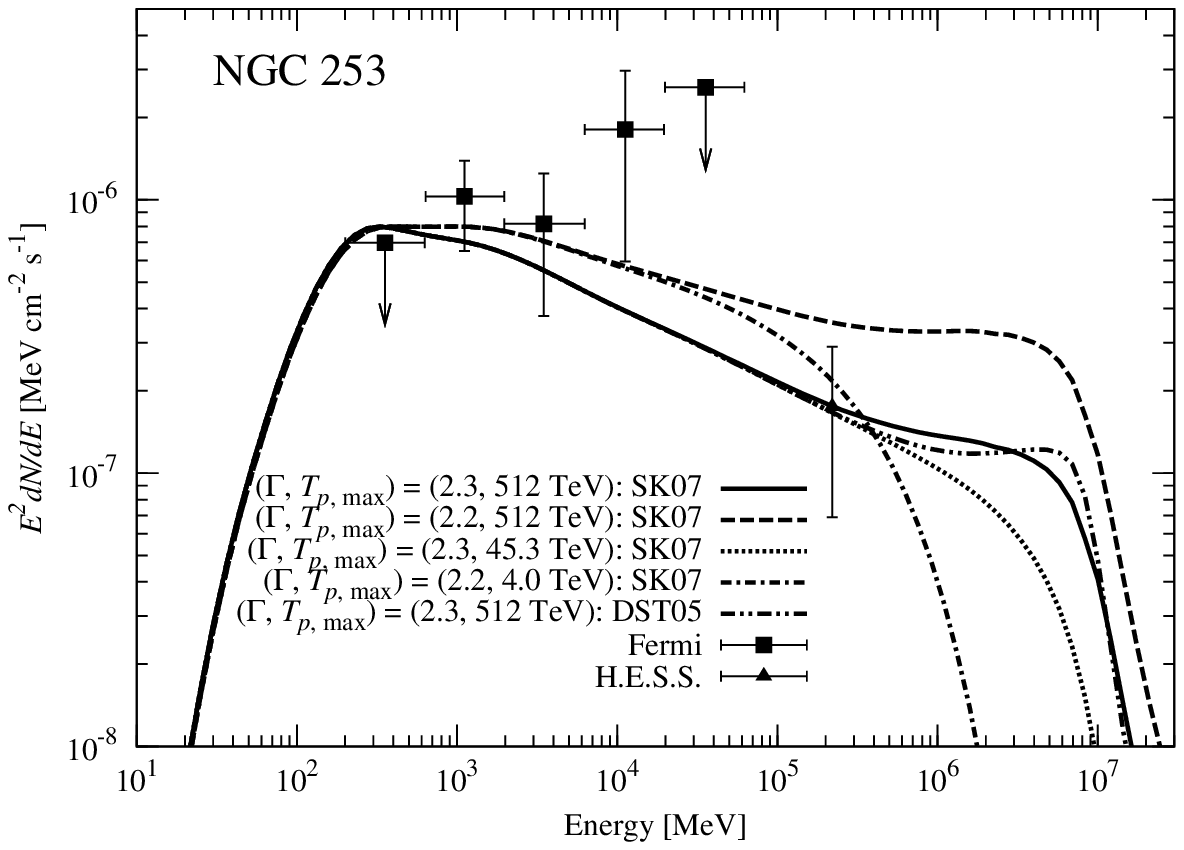}
\caption{Same as Fig. \ref{fig:f_sb}, but with the total flux for various $\Gamma$, $T_{p, \rm max}$, and interstellar radiation SED models. {\it Left} panel: Solid, dashed, dotted, and dot-dashed curve is the model for ($\Gamma, T_{p, \rm max}) =$ $(2.4, 512\  {\rm TeV})$, $(2.3, 512\  {\rm TeV})$, $(2.4, 45.3\  {\rm TeV})$, and $(2.4, 4.0\  {\rm TeV})$ with the SK07 interstellar SED model, respectively. Double dot-dashed curve is the model for ($\Gamma, T_{p, \rm max}) =$ $(2.4, 512\  {\rm TeV})$ with the dCdP09 interstellar SED model. {\it Right} panel: Solid, dashed, dotted, and dot-dashed curve is the model for ($\Gamma, T_{p, \rm max}) =$ $(2.3, 512\  {\rm TeV})$, $(2.2, 512\  {\rm TeV})$, $(2.3, 45.3\  {\rm TeV})$, and $(2.2, 4.0\  {\rm TeV})$ with the SK07 SED model, respectively. Double dot-dashed curve is the model for ($\Gamma, T_{p, \rm max}) =$ $(2.3, 512\  {\rm TeV})$ with the DST05 SED model. }
\label{fig:f_index_max}
\end{center}
\end{figure*}

Fig. \ref{fig:f_index_max} also shows the expected $\gamma$-ray spectra of these two starburst galaxies but changing the intrinsic spectral index of proton, $\Gamma$, the highest proton kinetic energy, $T_{p, \rm max}$, and interstellar radiation field SED models for comparison. In the case of M82, the total $\gamma$-ray flux above 1 TeV becomes 48 \% and 22 \% greater than the absorbed flux with ($\Gamma, T_{p, \rm max}) = (2.3, 512\  {\rm TeV})$ and $(2.4, 45.3 \ {\rm TeV})$ model, respectively.  When we use the dCdP09 interstellar radiation field SED model with the standard parameters, the total flux above 1 TeV is 47 \% greater than the absorbed flux. In the case of NGC 253, the total $\gamma$-ray flux above 1 TeV becomes 52 \% and 14 \% greater than the absorbed flux with ($\Gamma, T_{p, \rm max}) = (2.2, 512\  {\rm TeV})$ and $(2.3, 45.3 \ {\rm TeV})$ model, respectively.  When we use the DST05 SED model with the standard parameters, the total flux above 1 TeV is 37 \% greater than the absorbed flux. Therefore, even if we use other interstellar radiation field model which do not fit to the near infrared observed data, the effect of the cascade emission would not be changed significantly. This is because the near infrared photon density is not high enough to absorb VHE $\gamma$-rays effectively.

It would be difficult to explain the data by the standard parameters, but ($\Gamma, T_{p, \rm max}) = (2.2, 4.0\  {\rm TeV})$ model would be able to explain the data. We should note, however, note that, in this case, cascade emissions for this model would not contribute to the VHE spectrum since $>$TeV emission is weak. However, $T_{p, \rm max}=4.0$ TeV model for M82 is inconsistent with the data. We will need to wait for much longer time integration data from {\it Fermi} and future TeV $\gamma$-ray observatories to discuss this point in greater detail.

\section{Discussion}
\subsection{Implications for future VHE observations}
To see the feature of the cascade emission, it will be important to detect these two nearby starburst galaxies with a high signal-to-noise ratio with future VHE observations. The Cherenkov Telescope Array (CTA) is planned for the next generation imaging atmospheric Cherenkov telescope (IACT) \footnote{CTA: http://www.cta-observatory.org/}. The sensitivity and energy range of CTA will be improved by one order of magnitude compared to current IACTs such as H.E.S.S, MAGIC, and VERITAS. Here we investigate the required observational time to see the absorption and cascade signature with high significance by future CTA observations following the argument in \S 3 and \S 4 in \citet{ino10}.

First, the required observational time to detect the integral $\gamma$-ray flux above 1 TeV with 5 $\sigma$ is 1.3 hrs for M82 and 1.0 hr for NGC 253, respectively. Next, to see the cascade features in the spectrum, we set the required signal-to-noise ratio to be 5$\sigma$ per logarithmic energy bin width of $\Delta E/E=0.1$. This corresponds to a $19 \sigma$ detection for integrated flux. From the expected flux in our standard model for M82, the required observing time to achieve this signal-to-noise ratio is 14, 19, and 330 hours for 300 GeV, 1 TeV, and 3 TeV, respectively. In the case of NGC 253, the required time is 14, 14, and 180 hours for the above 3 energy bands in our standard model, respectively. Thus, it would be possible for CTA to obtain high energy resolution spectra to see the cascade effects within reasonable observational times.

\section{Conclusions}
In this paper, we estimated the contribution from the cascade emissions which are resulted from the inverse Compton scattering of the interstellar radiation field in the galaxies by the $e^+e^-$ pairs created by $\gamma$-ray attenuation. To evaluate the $\gamma$-ray attenuation, we first modeled the $\gamma-\gamma$ optical depth of very high energy $\gamma$-ray ($>$30 GeV) absorption by the interstellar radiation field via positron and electron, $e^+e^-$, pair creation in the two nearby bright starburst galaxies, M82 and NGC 253. To take account of the interstellar radiation, we adopted the SED model in the central star forming region provided by \citet{sie07}. Since $\gamma$-rays are created inside of the central star formation region, absorption effects from such obscured emission is important. We found that the attenuation effects becomes significant above $\sim 5$ TeV and $\sim9$ TeV for M82 and NGC 253, respectively. 

By using the interstellar radiation field SED model by  \citet{sie07} and our $\gamma-\gamma$ optical depth model based on it, we found that the cascade emissions would be a probe of the existence of high energy cosmic-rays ($>10$ TeV) in starburst galaxies. The resulting total flux including cascade emission would be $\sim$18 \% and $\sim$45\% higher than the absorbed flux above 1 TeV with the maximum proton kinetic energy, 45.3 TeV and 512 TeV, respectively. Even when we use the \citet{dcdp09} interstellar SED model for M82 and the \citet{ds05} interstellar SED model for NGC 253, the effect of cascade emission is not significantly changed. These differences would be a new indirect investigator of the highest cosmic-ray energy, although it would be difficult to see the direct $\gamma$-ray signature because of the attenuation. It might be possible to investigate this cascade signature through a detailed differential spectrum observation with future high signal-to-noise ratio $\gamma$-ray observations such as CTA by comparing with the theoretical $\gamma$-ray emission models including this cascade effect.

\acknowledgments
The author would like to thank M. Hayashida, K. Matsubayashi, M. Sawada, and T. Totani for useful discussions; L. Herman and T. Kamae for providing their numerical code; and A. R. Jenner for his careful reading of the draft. The author also thanks the anonymous referee comments that improved this paper. This work was supported by the Grant-in-Aid for the Global COE Program "The Next Generation of Physics, Spun from Universality and Emergence" from the Ministry of Education, Culture, Sports, Science and Technology (MEXT) of Japan. The author acknowledges support by the Research Fellowship of the Japan Society for the Promotion of Science (JSPS).


\begin{thebibliography}{61}
\expandafter\ifx\csname natexlab\endcsname\relax\def\natexlab#1{#1}\fi

\bibitem[{{Aaronson}(1977)}]{aar77}
{Aaronson}, M. 1977, PhD thesis, Harvard Univ., Cambridge, MA.

\bibitem[{{Abdo} {et~al.}(2010{\natexlab{a}})}]{abd10_sb}
{Abdo}, A.~A. {et~al.} 2010{\natexlab{a}}, \apjl, 709, L152

\bibitem[{{Abdo} {et~al.}(2010{\natexlab{b}})}]{abd10_casa}
---. 2010{\natexlab{b}}, \apjl, 710, L92

\bibitem[{{Abdo} {et~al.}(2010{\natexlab{c}})}]{abd10_lmc}
---. 2010{\natexlab{c}}, \aap, 512, A7+

\bibitem[{{Acero} {et~al.}(2009)}]{ace09}
{Acero}, F. {et~al.} 2009, Science, 326, 1080

\bibitem[{{Aharonian} {et~al.}(1994){Aharonian}, {Coppi}, \& {Voelk}}]{aha94}
{Aharonian}, F.~A., {Coppi}, P.~S., \& {Voelk}, H.~J. 1994, \apjl, 423, L5

\bibitem[{{Aharonian} {et~al.}(2004)}]{aha04}
{Aharonian}, F.~A. {et~al.} 2004, \nat, 432, 75

\bibitem[{{Akyuz} {et~al.}(1991){Akyuz}, {Brouillet}, \& {Ozel}}]{aky91}
{Akyuz}, A., {Brouillet}, N., \& {Ozel}, M.~E. 1991, \aap, 248, 419

\bibitem[{{Ando}(2004)}]{and04}
{Ando}, S. 2004, \mnras, 354, 414

\bibitem[{{Blumenthal} \& {Gould}(1970)}]{blu70}
{Blumenthal}, G.~R. \& {Gould}, R.~J. 1970, Reviews of Modern Physics, 42, 237

\bibitem[{{Chini} {et~al.}(1984){Chini}, {Kreysa}, {Mezger}, \&
  {Gemuend}}]{chi84}
{Chini}, R., {Kreysa}, E., {Mezger}, P.~G., \& {Gemuend}, H. 1984, \aap, 137,
  117

\bibitem[{{Dai} {et~al.}(2002){Dai}, {Zhang}, {Gou}, {M{\'e}sz{\'a}ros}, \&
  {Waxman}}]{dai02}
{Dai}, Z.~G., {Zhang}, B., {Gou}, L.~J., {M{\'e}sz{\'a}ros}, P., \& {Waxman},
  E. 2002, \apjl, 580, L7

\bibitem[{{de Cea del Pozo} {et~al.}(2009){de Cea del Pozo}, {Torres}, \&
  {Rodriguez Marrero}}]{dcdp09}
{de Cea del Pozo}, E., {Torres}, D.~F., \& {Rodriguez Marrero}, A.~Y. 2009,
  \apj, 698, 1054

\bibitem[{{Domingo-Santamar{\'{\i}}a} \& {Torres}(2005)}]{ds05}
{Domingo-Santamar{\'{\i}}a}, E. \& {Torres}, D.~F. 2005, \aap, 444, 403

\bibitem[{{Elias} {et~al.}(1978)}]{eli78}
{Elias}, J.~H. {et~al.} 1978, \apj, 220, 25

\bibitem[{{F{\"o}rster Schreiber} {et~al.}(2003){F{\"o}rster Schreiber},
  {Sauvage}, {Charmandaris}, {Laurent}, {Gallais}, {Mirabel}, \&
  {Vigroux}}]{for03}
{F{\"o}rster Schreiber}, N.~M., {Sauvage}, M., {Charmandaris}, V., {Laurent},
  O., {Gallais}, P., {Mirabel}, I.~F., \& {Vigroux}, L. 2003, \aap, 399, 833

\bibitem[{{Freedman} {et~al.}(1994)}]{fre94}
{Freedman}, W.~L. {et~al.} 1994, \apj, 427, 628

\bibitem[{{Ginzburg} \& {Syrovatskii}(1964)}]{gin64}
{Ginzburg}, V.~L. \& {Syrovatskii}, S.~I. 1964, {The Origin of Cosmic Rays},
  ed. {Ginzburg, V.~L.~\& Syrovatskii, S.~I.}

\bibitem[{{Gould} \& {Schr{\'e}der}(1966)}]{gou66}
{Gould}, R.~J. \& {Schr{\'e}der}, G. 1966, Physical Review Letters, 16, 252

\bibitem[{{Hayakawa}(1969)}]{hay69}
{Hayakawa}, S. 1969, {Cosmic ray physics. Nuclear and astrophysical aspects},
  ed. {Hayakawa, S.}

\bibitem[{{Hildebrand} {et~al.}(1977){Hildebrand}, {Whitcomb}, {Winston},
  {Stiening}, {Harper}, \& {Moseley}}]{hil77}
{Hildebrand}, R.~H., {Whitcomb}, S.~E., {Winston}, R., {Stiening}, R.~F.,
  {Harper}, D.~A., \& {Moseley}, S.~H. 1977, \apj, 216, 698

\bibitem[{{Hughes} {et~al.}(1994){Hughes}, {Gear}, \& {Robson}}]{hug94}
{Hughes}, D.~H., {Gear}, W.~K., \& {Robson}, E.~I. 1994, \mnras, 270, 641

\bibitem[{{Hughes} {et~al.}(1990){Hughes}, {Robson}, \& {Gear}}]{hug90}
{Hughes}, D.~H., {Robson}, E.~I., \& {Gear}, W.~K. 1990, \mnras, 244, 759

\bibitem[{{Inoue} \& {Totani}(2009)}]{ino09}
{Inoue}, Y. \& {Totani}, T. 2009, \apj, 702, 523

\bibitem[{{Inoue} {et~al.}(2010){Inoue}, {Totani}, \& {Mori}}]{ino10}
{Inoue}, Y., {Totani}, T., \& {Mori}, M. 2010, \pasj, 62, 1005

\bibitem[{{Jaffe} {et~al.}(1984){Jaffe}, {Becklin}, \& {Hildebrand}}]{jaf84}
{Jaffe}, D.~T., {Becklin}, E.~E., \& {Hildebrand}, R.~H. 1984, \apjl, 285, L31

\bibitem[{{Jarrett} {et~al.}(2003){Jarrett}, {Chester}, {Cutri}, {Schneider},
  \& {Huchra}}]{jar03}
{Jarrett}, T.~H., {Chester}, T., {Cutri}, R., {Schneider}, S.~E., \& {Huchra},
  J.~P. 2003, \aj, 125, 525

\bibitem[{{Jelley}(1966)}]{jel66}
{Jelley}, J.~V. 1966, Physical Review Letters, 16, 479

\bibitem[{{Johnson}(1966)}]{joh66}
{Johnson}, H.~L. 1966, \apj, 143, 187

\bibitem[{{Kamae} {et~al.}(2005){Kamae}, {Abe}, \& {Koi}}]{kam05}
{Kamae}, T., {Abe}, T., \& {Koi}, T. 2005, \apj, 620, 244

\bibitem[{{Kamae} {et~al.}(2006){Kamae}, {Karlsson}, {Mizuno}, {Abe}, \&
  {Koi}}]{kam06}
{Kamae}, T., {Karlsson}, N., {Mizuno}, T., {Abe}, T., \& {Koi}, T. 2006, \apj,
  647, 692

\bibitem[{{Karachentsev} {et~al.}(2003)}]{kar03}
{Karachentsev}, I.~D. {et~al.} 2003, \aap, 404, 93

\bibitem[{{Karlsson} \& {Kamae}(2008)}]{kar08}
{Karlsson}, N. \& {Kamae}, T. 2008, \apj, 674, 278

\bibitem[{{Klein} {et~al.}(1988){Klein}, {Wielebinski}, \& {Morsi}}]{kle88}
{Klein}, U., {Wielebinski}, R., \& {Morsi}, H.~W. 1988, \aap, 190, 41

\bibitem[{{Kleinmann} \& {Low}(1970)}]{kle70}
{Kleinmann}, D.~E. \& {Low}, F.~J. 1970, \apjl, 159, L165+

\bibitem[{{Kneiske} \& {Mannheim}(2008)}]{kne08}
{Kneiske}, T.~M. \& {Mannheim}, K. 2008, \aap, 479, 41

\bibitem[{{Koyama} {et~al.}(1995){Koyama}, {Petre}, {Gotthelf}, {Hwang},
  {Matsuura}, {Ozaki}, \& {Holt}}]{koy95}
{Koyama}, K., {Petre}, R., {Gotthelf}, E.~V., {Hwang}, U., {Matsuura}, M.,
  {Ozaki}, M., \& {Holt}, S.~S. 1995, \nat, 378, 255

\bibitem[{{Krugel} {et~al.}(1990){Krugel}, {Chini}, {Klein}, {Lemke},
  {Wielebinski}, \& {Zylka}}]{kru90}
{Krugel}, E., {Chini}, R., {Klein}, U., {Lemke}, R., {Wielebinski}, R., \&
  {Zylka}, R. 1990, \aap, 240, 232

\bibitem[{{Mayya} {et~al.}(2006){Mayya}, {Bressan}, {Carrasco}, \&
  {Hernandez-Martinez}}]{may06}
{Mayya}, Y.~D., {Bressan}, A., {Carrasco}, L., \& {Hernandez-Martinez}, L.
  2006, \apj, 649, 172

\bibitem[{{Melo} {et~al.}(2002){Melo}, {P{\'e}rez Garc{\'{\i}}a},
  {Acosta-Pulido}, {Mu{\~n}oz-Tu{\~n}{\'o}n}, \& {Rodr{\'{\i}}guez
  Espinosa}}]{mel02}
{Melo}, V.~P., {P{\'e}rez Garc{\'{\i}}a}, A.~M., {Acosta-Pulido}, J.~A.,
  {Mu{\~n}oz-Tu{\~n}{\'o}n}, C., \& {Rodr{\'{\i}}guez Espinosa}, J.~M. 2002,
  \apj, 574, 709

\bibitem[{{Murase} {et~al.}(2007){Murase}, {Asano}, \& {Nagataki}}]{mur07}
{Murase}, K., {Asano}, K., \& {Nagataki}, S. 2007, \apj, 671, 1886

\bibitem[{{Nagano} \& {Watson}(2000)}]{nag00}
{Nagano}, M. \& {Watson}, A.~A. 2000, Reviews of Modern Physics, 72, 689

\bibitem[{{Paglione} {et~al.}(1996){Paglione}, {Marscher}, {Jackson}, \&
  {Bertsch}}]{pag96}
{Paglione}, T.~A.~D., {Marscher}, A.~P., {Jackson}, J.~M., \& {Bertsch}, D.~L.
  1996, \apj, 460, 295

\bibitem[{{Persic} {et~al.}(2008){Persic}, {Rephaeli}, \& {Arieli}}]{per08}
{Persic}, M., {Rephaeli}, Y., \& {Arieli}, Y. 2008, \aap, 486, 143

\bibitem[{{Radovich} {et~al.}(2001){Radovich}, {Kahanp{\"a}{\"a}}, \&
  {Lemke}}]{rad01}
{Radovich}, M., {Kahanp{\"a}{\"a}}, J., \& {Lemke}, D. 2001, \aap, 377, 73

\bibitem[{{Razzaque} {et~al.}(2004){Razzaque}, {M{\'e}sz{\'a}ros}, \&
  {Zhang}}]{raz04}
{Razzaque}, S., {M{\'e}sz{\'a}ros}, P., \& {Zhang}, B. 2004, \apj, 613, 1072

\bibitem[{{Rephaeli} {et~al.}(2010){Rephaeli}, {Arieli}, \& {Persic}}]{rep10}
{Rephaeli}, Y., {Arieli}, Y., \& {Persic}, M. 2010, \mnras, 401, 473

\bibitem[{{Rieke} {et~al.}(1973){Rieke}, {Harper}, {Low}, \&
  {Armstrong}}]{rie73}
{Rieke}, G.~H., {Harper}, D.~A., {Low}, F.~J., \& {Armstrong}, K.~R. 1973,
  \apjl, 183, L67+

\bibitem[{{Rieke} {et~al.}(1980){Rieke}, {Lebofsky}, {Thompson}, {Low}, \&
  {Tokunaga}}]{rie80}
{Rieke}, G.~H., {Lebofsky}, M.~J., {Thompson}, R.~I., {Low}, F.~J., \&
  {Tokunaga}, A.~T. 1980, \apj, 238, 24

\bibitem[{{Rieke} \& {Low}(1972)}]{rie72}
{Rieke}, G.~H. \& {Low}, F.~J. 1972, \apjl, 176, L95+

\bibitem[{{Rieke} \& {Low}(1975)}]{rie75}
---. 1975, \apj, 197, 17

\bibitem[{{Siebenmorgen} \& {Kr{\"u}gel}(2007)}]{sie07}
{Siebenmorgen}, R. \& {Kr{\"u}gel}, E. 2007, \aap, 461, 445

\bibitem[{{Telesco} \& {Gezari}(1992)}]{tel92}
{Telesco}, C.~M. \& {Gezari}, D.~Y. 1992, \apj, 395, 461

\bibitem[{{Telesco} \& {Harper}(1980)}]{tel80}
{Telesco}, C.~M. \& {Harper}, D.~A. 1980, \apj, 235, 392

\bibitem[{{Thompson} {et~al.}(2007){Thompson}, {Quataert}, \& {Waxman}}]{tho07}
{Thompson}, T.~A., {Quataert}, E., \& {Waxman}, E. 2007, \apj, 654, 219

\bibitem[{{Torres}(2004)}]{tor04}
{Torres}, D.~F. 2004, \apj, 617, 966

\bibitem[{{Ulvestad}(2000)}]{ulv00}
{Ulvestad}, J.~S. 2000, \aj, 120, 278

\bibitem[{{Venters}(2010)}]{ven10}
{Venters}, T.~M. 2010, \apj, 710, 1530

\bibitem[{{VERITAS Collaboration} {et~al.}(2009)}]{ver09}
{VERITAS Collaboration} {et~al.} 2009, \nat, 462, 770

\bibitem[{{Voelk} {et~al.}(1989){Voelk}, {Klein}, \& {Wielebinski}}]{voe89}
{Voelk}, H.~J., {Klein}, U., \& {Wielebinski}, R. 1989, \aap, 213, L12

\bibitem[{{Wang} {et~al.}(2001){Wang}, {Dai}, \& {Lu}}]{wan01}
{Wang}, X.~Y., {Dai}, Z.~G., \& {Lu}, T. 2001, \apj, 556, 1010

\end{thebibliography}
\end{document}